\documentclass[12pt]{iopart}

\usepackage[backend=biber,style=chem-angew,sorting=none]{biblatex}
\usepackage{hyperref}
\usepackage{caption}
\usepackage[dvipsnames]{xcolor}
\usepackage{xspace}
\usepackage{array}

\usepackage{tikz}
\usetikzlibrary{decorations.pathmorphing}
\tikzstyle{mybox}=[draw=black,fill=white,very thick,rectangle,rounded corners,inner sep=20pt,inner ysep=10pt]
\tikzstyle{fancytitle}=[fill=black,text=white,font=\bfseries]
\usepackage{fancyvrb}
\VerbatimFootnotes

\usepackage{etoolbox}
\makeatletter
\newcommand{\mainmatter}{%
  \setcounter{footnote}{0}%
  \patchcmd{\@makefntext}{\fnsymbol}{\alph}{}{}
  \patchcmd{\@thefnmark}{\fnsymbol}{\alph}{}{}
  \def\@makefnmark{\textsuperscript{\alph{footnote}}}
}
\makeatother

\newcolumntype{C}[1]{>{\centering\let\newline\\\arraybackslash\hspace{0pt}}m{#1}}

\usepackage[framemethod=TikZ]{mdframed}
\usepackage{float}
\usepackage[scaled]{beramono}
\usepackage[T1]{fontenc}
\usepackage{textcomp}
\usepackage{listings}

\lstnewenvironment{boxed}[1][]{
    \lstset{basicstyle=\ttfamily\scriptsize,breaklines=true,upquote=true,numbers=left,moredelim=**[is][\color{blue}]{@}{@}}
    \mdframed[outerlinewidth=.75pt,roundcorner=10pt,innerleftmargin=25pt,nobreak=true]
}{\endmdframed}
\lstnewenvironment{iboxed}[1][]{
    \lstset{basicstyle=\ttfamily\scriptsize,breaklines=true,upquote=true,moredelim=**[is][\color{red}]{@}{@},moredelim=**[is][\color{OliveGreen}]{`}{`},moredelim=**[is][\color{LimeGreen}]{!}{!}}
    \mdframed[outerlinewidth=.75pt,roundcorner=10pt,innerleftmargin=25pt,nobreak=true]
}{\endmdframed}
\lstnewenvironment{titleboxed}[1][]{
    \lstset{basicstyle=\ttfamily\scriptsize,breaklines=true,upquote=true}
}{}

\newcommand{\beq}{\begin{equation}}
\newcommand{\eeq}{\end{equation}}
\newcommand{\beqn}{\begin{eqnarray}}
\newcommand{\eeqn}{\end{eqnarray}}

\newcommand{\nrpylatex}{{\texttt{NRPyLaTeX}}\xspace}
\newcommand{\nl}{{\texttt{NL}}\xspace}
\newcommand{\ipython}{{\texttt{IPython}}\xspace}
\newcommand{\jupyter}{{\texttt{Jupyter}}\xspace}
\newcommand{\python}{{\texttt{Python}}\xspace}

\newcommand{\ccpp}{{\texttt{C/C++}}\xspace}
\newcommand{\sympy}{{\texttt{SymPy}}\xspace}

\newcommand{\einsteinpy}{{\texttt{EinsteinPy}}\xspace}
\newcommand{\gravipy}{{\texttt{GraviPy}}\xspace}
\newcommand{\xtensor}{{\texttt{xTensor}}\xspace}
\newcommand{\mathematica}{{\texttt{Mathematica}}\xspace}
\newcommand{\maple}{{\texttt{Maple}}\xspace}
\newcommand{\cadabra}{{\texttt{Cadabra}}\xspace}
\newcommand{\nrpy}{{\texttt{NRPy+}}\xspace}
\newcommand{\git}{{\texttt{git}}\xspace}

\newcommand{\latex}{{\LaTeX}\xspace}

\def\eadnew#1#2{\address{#2 E-mail: \mailto{#1}}}

\begin{document}
\date{\today}
\title{\texttt{NRPyLaTeX}: A LaTeX interface to computer algebra systems for general relativity}
\author{
  Kenneth J.~Sible$^{1,\dagger}$,
  Zachariah B.~Etienne$^{2,3,4,\ddagger}$}

\address{$^{1}$ Department of Computer Science \& Engineering, University of Notre Dame, Notre Dame, IN 46556, USA}
\address{$^{2}$ Department of Physics, University of Idaho, Moscow, ID 83843, USA}
\address{$^{3}$ Department of Physics \& Astronomy, West Virginia University, Morgantown, WV 26506, USA}
\address{$^{4}$ Center for Gravitational Waves and Cosmology, West Virginia University, Chestnut Ridge Research Building, Morgantown, WV 26505, USA}
\eadnew{ksible@nd.edu}{$^{\dagger}$}
\eadnew{zetienne@uidaho.edu}{$^{\ddagger}$}

\begin{abstract}
While each computer algebra system (CAS) contains its own unique syntax for inputting mathematical expressions, \latex is perhaps {\it the} most widespread language for typesetting mathematics. \nrpylatex (\nl) enables direct \latex input of complex tensorial expressions (written in Einstein notation) relevant to general relativity and differential geometry into the \sympy CAS. As \sympy also supports output compatible with the \mathematica and \maple CASs, \nl lowers the learning curve for inputting and manipulating tensorial expressions in three widely used CASs. 
\latex however is a typesetting language, and as such is not designed to resolve ambiguities in mathematical expressions. To address this, \nl implements a convenient configuration interface that, e.g., defines variables with certain  attributes. Configuration commands appear as \latex comments, so that entire \nl workflows can fit seamlessly into the \latex source code of scientific papers without interfering with the rendered mathematical expressions. Further, \nl adopts \nrpy's rigid syntax for indexed symbols (e.g., tensors), which enables \nl output to be directly converted into highly optimized \ccpp-code kernels using \nrpy. Finally \nl has robust and user-friendly error-handling, which catches common tensor indexing errors and reports unresolved ambiguities, further expediting the input and validation of \latex expressions into a CAS.
\end{abstract}


\submitto{\CQG}

\maketitle

\mainmatter

\section{Introduction}

\latex is perhaps {\it the} canonical markup language for communicating mathematical ideas in scientific papers, and most practitioners of general relativity and differential geometry are proficient in it. Further, computer algebra systems (CASs) like \sympy~\cite{meurer2017sympy}, \mathematica~\cite{Mathematica}, or \maple~\cite{Maple} are often an essential resource in these fields, but to our knowledge there are currently no CASs or CAS add-on packages, except for \cadabra~\cite{Peeters2018}, that support the input of tensorial/indexed \latex expressions adopting Einstein notation. Thus new practitioners are often required to learn one language for communicating mathematical ideas in scientific papers, and another to input \& manipulate mathematical expressions in a CAS. As a result expressions developed within a CAS often need to be translated into \latex or vice-versa when constructing a scientific paper, and human errors in transcription can become a menace to productivity and reproducibility of scientific results.

Various packages have been developed to ease the input of tensorial/indexed expressions using Einstein notation into CASs. For example, \xtensor~\cite{xtensor} is an open-source add-on package for the \mathematica CAS, greatly extending its capabilities for performing abstract tensor calculus. In short, \xtensor enables the definition and manipulation of tensors with arbitrary symmetries on multiple manifolds. Upon declaring a metric for a given manifold, covariant, Lie, and other derivatives may be computed, and associated curvature tensors may be immediately defined. As it is an extremely powerful add-on to one of the most popular and powerful CASs, \xtensor finds wide use across the community. Despite its success, perhaps the largest barrier to entry for new users is its dependency on the closed-source \mathematica, whose software licenses carry a price that can be too high for some users.


\cadabra is a powerful CAS for solving problems in field theory, with a similar design philosophy to \nrpy/\nrpylatex. Both \cadabra and \nrpy/\nrpylatex use the same three language system: \latex for mathematics, \python for programming, and \ccpp for computation. In contrast, most CASs, e.g. \sympy, use a single language for each of those respective tasks. However, \cadabra and \nrpy differ in three main respects. First, \nrpylatex is intended for general relativity. As such, \nrpylatex makes certain convenient inferences about indexing, consistent with common conventions in the general relativity and numerical relativity literature. Specifically, Latin and Greek indices default to 3D and 4D, respectively. In addition, \nrpylatex adopts notations common to the field of general relativity for specifying covariant (e.g., $D_i$ or $\nabla_i$) and Lie (e.g., $\mathcal{L}_{\beta}$) derivative operators. \cadabra, on the other hand, is intended for more general field theory use, and thus requires more user input to remove ambiguities. Second, as detailed below \nrpylatex is natively compatible with numerical workflows through \nrpy's C-code generation capabilities, as it adopts rigid syntax for naming indexed symbols (e.g., $v^i\equiv$\ \verb|vU[i]|) and specifying derivatives that are to be computed numerically. In contrast, \cadabra is largely geared toward symbolic workflows. Third, \nrpylatex configurations (i.e., commands that users must input to remove ambiguities in the symbolic expressions, like specifying the symmetry properties of a tensor) exist as \LaTeX\ comments, so \nrpylatex workflows can be embedded completely inside of the source code of a \LaTeX\ document, eliminating the step of sharing these workflows through a separate document. As described below, this feature is particularly convenient in the context of scientific paper preprints on e.g., the \verb|arXiv|, which are usually downloadable as \LaTeX\ source. \cadabra's configuration, on the other hand, does not appear as \LaTeX\ comments, making it less convenient to share configurations.



\sympy is another very powerful CAS used widely across the scientific
community. Unlike \mathematica or \maple, \sympy is built upon \python
and is open-source (under the permissive 3-Clause BSD license). While
\sympy contains both basic support for tensors and differential
geometry, as well as basic, experimental \latex input capabilities,
\latex input of tensorial/indexed expressions is not currently
supported\footnote{as of \sympy 1.9}. Instead, to input and manipulate
tensorial/indexed expressions found in general relativity and
differential geometry contexts, a few add-on packages for \sympy have
been developed. These packages are somewhat analogous to \xtensor, but
they benefit from the fact that \sympy possesses no expensive software
license. \einsteinpy~\cite{einsteinpy} is one such (open-source, free)
package, which extends basic tensor support in \sympy to handle
metrics and construction of their associated connections and curvature
tensors (Riemann \& Ricci). It also contains friendly \jupyter
notebook documentation and functionality for e.g., plotting geodesics
in various spacetimes. The \gravipy~\cite{gravipy} add-on package to
\sympy possesses similar basic tensor symbolic manipulation functionality as \einsteinpy, but in addition e.g., supports the construction of the Einstein tensor, as well as user-defined tensors and covariant derivatives.

While \xtensor, \einsteinpy, and \gravipy are exemplary packages that effectively convert popular CASs into powerful tools for general r\textit{}elativity and differential geometry, each depends on its own custom, non-\LaTeX\ syntax for inputting tensorial expressions. As discussed, this adds a slight learning curve for new users, and opens the door to \latex$\leftrightarrow$~CAS transcription errors when preparing scientific papers.

Our new, open-source\footnote{under the permissive 2-Clause BSD license} \nrpylatex package provides a multifaceted \latex interface to \sympy, with general relativity and differential geometry workloads in mind. \nrpylatex enables the input of tensorial expressions exactly as they would appear in the \latex source code of scientific papers, and it possesses native support for e.g., multiple metrics; custom tensors of arbitrary rank \& dimension; index raising \& lowering; and covariant, Lie, \& partial derivatives.

As \latex is a typesetting markup language, it is neither designed to resolve ambiguities in mathematical expressions, nor able to point out e.g., errors in Einstein notation. To address ambiguities (e.g., which rank-2 tensor represents the metric), \nrpylatex contains a robust and user-friendly configuration interface, which enables one to define variables, assign attributes, ignore substrings, and perform (syntactic) string replacement. For example, this includes the ability to declare variables, define tensors, input into \sympy un-rendered equations, declare arbitrary operators, and provide aliases for variable names. Conveniently this configuration interface exists entirely within \latex comments, enabling it to be interspersed with the rendered \latex without itself being rendered. Thus it provides a means to share CAS worksheets {\it within the source code of scientific papers}, while minimizing the chances of a transcription error when translating CAS input expressions to a \latex-ed scientific paper. It also provides a new means for inputting \latex-ed equations (found in e.g., scientific paper source codes on the arXiv) directly into CASs.

To address errors in Einstein notation (e.g., ``indexing'' errors) and to help guide the user in resolving ambiguities in mathematical expressions, \nrpylatex contains a comprehensive exception handling infrastructure. For example, this infrastructure will return errors when a \latex-ed tensorial expression written in Einstein notation contains a repeated ``down'' index, or an unbalanced free index across the equality, or the use of a covariant derivative without specifying the associated metric. 

As its name implies, \nrpylatex falls under the umbrella of our \sympy-based \nrpy\footnote{\nrpy is short for ``Python-based code generation for numerical relativity... and beyond!''.} framework~\cite{Ruchlin:2017com,NRPy_website}; \nrpylatex can be used seamlessly for not only symbolic but also numerical workflows using \nrpy. In short, \nrpy builds upon and extends \sympy's native capabilities, to convert often complex indexed expressions (e.g., tensorial expressions) directly into highly optimized \ccpp-code kernels with common-subexpression elimination; single-instruction, multiple-data (SIMD); and finite-difference derivative code generation capabilities. \ccpp-code kernels like these form the heart of numerical relativity codes.

The remainder of the paper is organized as follows. In Sec.~\ref{sec:method} we discuss the basic features and internal structure of \nrpylatex; and in Sec.~\ref{sec:examples} present symbolic and numerical workflows, as well as usage and exception handling. Finally in Sec.~\ref{sec:conclusions} we summarize the basic results and present plans for future work.

\section{Basic Features and Methodology}
\label{sec:method}

\nrpylatex provides a \latex frontend to the \sympy CAS\footnote{\sympy provides a pipeline from \nrpylatex to \mathematica/\maple through their Printing module.}, supporting the input of tensorial expressions written in Einstein notation. It includes a number of features with general relativity and differential geometry applications in mind, including covariant and Lie derivatives; spacetime and reference metrics; Levi-Civita and Christoffel symbols; as well as metric inverses and determinants, just to name a few.

At its core the \nrpylatex frontend is a lexer/parser\footnote{\nrpylatex tokenizes the input \latex, using substring pattern matching, and parses the resulting token iterator at a syntactic level, including proper handling of Einstein notation.} with a powerful configuration interface that addresses ambiguities (e.g., specifying which metric is associated with a covariant derivative, or the dimension of a tensor) and defines arbitrary variable/operator macros. Configuration options are input as \latex comments so that all information needed to construct symbolic expressions can be shared in, e.g., a scientific paper's \latex source code.

\nrpylatex adopts the \nrpy notation for indexed symbols, so that e.g., $g_{\mu\nu}$ is accessible as a nested list of \sympy symbols \verb!gDD[mu][nu]!, where e.g., $g_{tt}$ is stored in \verb|gDD[0][0]|. Here \verb!D! denotes a covariant ({\it down}) index. Similarly \verb!U! is appended to the tensor name for each contravariant ({\it up}) index. In fact, for each indexed symbol of dimension $n$ and rank $r$, \nrpylatex will create an $r$-dimensional array (or symbol) of size $n\times \ldots \times n = n^r$ using \nrpy notation.

As \nrpy is designed to convert \sympy expressions into highly optimized \ccpp-code kernels, compatibility with \sympy and \nrpy enables \nrpylatex to interface well with symbolic and numerical workloads, respectively. For example, specifying that the partial derivative $g_{\mu\nu,\delta}$ should be evaluated numerically requires that one simply first define the variable \verb!gDD! with a placeholder differentiation type in a \nrpylatex configuration line. Once this is completed, all symbolic first derivatives of $g_{\mu\nu}$ will be accessed in the \nrpy-compatible syntax \verb!gDD_dD[mu][nu][delta]!, second derivatives as \verb!gDD_dDD[mu][nu][delta][gamma]!, and so forth. \nrpy can interpret such partial derivative symbols as requiring numerical (finite-difference) differentiation and generate the appropriate \ccpp-code kernel.\footnote{\nrpylatex extends the \nrpy notation for differentiation, in order to support Lie derivatives with respect to an arbitrary vector $v^i$ \verb+_ld[v]D..D+ and covariant derivatives (\verb+_cdD+, \verb+_cdU+, \verb+_cdDU+, etc.)}

Finally, \nrpylatex is developed using modern software engineering techniques, including \git version control\footnote{\nrpylatex source code can be downloaded from \url{https://github.com/zachetienne/nrpylatex}, though it is generally advisable to install \nrpylatex through \texttt{pip} via \texttt{pip install nrpylatex}\ .}, robust exception handling, interactive \jupyter notebook documentation, convenient \ipython/\jupyter interfaces, and more than 80 code-validation unit tests. These tests check \nrpylatex's exception handling, as well as a large number of challenging tensor parsing edge cases and use case examples. Perhaps the most sophisticated such use case is \nrpylatex's ability to reproduce the entire covariant BSSN~\cite{BrownCovariantBSSN,Montero:2012yr,Baumgarte:2012xy,Ruchlin:2017com} system of equations in a \nrpy-compatible form; the generated symbolic expressions have been validated against the handwritten \sympy expressions for the BSSN formalism appearing within \nrpy. Associated with the BSSN validation test is a \jupyter notebook tutorial that demonstrates precisely how to input such equations into \nrpylatex. All validation tests are run within a continuous integration infrastructure, which is triggered upon any commit to the \git repository. Further \nrpylatex exists within the \python Package Index (PyPI) and thus can be installed from a standard \python installation via \verb!pip install nrpylatex!. Though, for the purpose of reproducing results in this paper, one should specify version \verb!1.1.1!.

Next we demonstrate \nrpylatex's basic features through real-world examples, highlighting both its ability to handle symbolic (Sec.~\ref{ex:schwarzschild}) and numerical/\nrpy (Sec.~\ref{ex:hamiltonian}) workloads, as well as its exception handling (Sec.~\ref{ex:exceptionhandling}).

\section{Examples}
\label{sec:examples}

\subsection{Symbolic Workload Example: Schwarzschild Solution to Einstein's Equations}
\label{ex:schwarzschild}

\begin{figure}[H]
\vspace{-0.4cm}
\begin{boxed}
\begin{align}
    % attrib coord [t, r, \theta, \phi]
    % vardef -const G, M
    % vardef -zero gDD::4D  @%% initialize every component of gDD to zero@
    g_{t t} &= -\left(1 - \frac{2GM}{r}\right) \\ @%% g_{t t} = g_{0 0}@
    g_{r r} &=  \left(1 - \frac{2GM}{r}\right)^{-1} \\
    g_{\theta\theta} &= r^2 \\
    g_{\phi\phi} &= r^2 \sin^2\theta \\
    % assign -metric gDD  @%% inverse gUU, determinant det(gDD), and connection GammaUDD@
    R^\alpha{}_{\beta\mu\nu} &= \partial_\mu \Gamma^\alpha_{\beta\nu} - \partial_\nu \Gamma^\alpha_{\beta\mu} + \Gamma^\alpha_{\mu\gamma} \Gamma^\gamma_{\beta\nu} - \Gamma^\alpha_{\nu\sigma} \Gamma^\sigma_{\beta\mu} \\
    R_{\beta\nu} &= R^\alpha{}_{\beta\alpha\nu} \\
    R &= g^{\beta\nu} R_{\beta\nu} \\
    G_{\beta\nu} &= R_{\beta\nu} - \frac{R}{2} g_{\beta\nu} \\
    K &= R^{\alpha\beta\mu\nu} R_{\alpha\beta\mu\nu} @%% automatic index raising & lowering@
\end{align}
\end{boxed}
\vspace*{-\baselineskip}
\captionof{figure}{Symbolic workload example: Schwarzschild solution of general relativity}
\label{fig:schwarzschild}
\end{figure}

In this example, we demonstrate \nrpylatex's ability to handle symbolic workloads, by first defining the Schwarzschild metric, and then constructing the corresponding Riemann tensor $R^\alpha{}_{\beta\mu\nu}$, Ricci tensor $R_{\beta\nu}$, Ricci scalar $R$, Einstein tensor $G_{\beta\nu}$, and the Kretschmann scalar $K$. As shown in Fig.~\ref{fig:schwarzschild}, Lns.~2--3, the coordinate system and all constants (ignored during differentiation) must first be specified in the \nrpylatex configuration. Then in Ln.~4, the 4-metric $g_{\mu\nu}$ is initialized to zero (this ensures off-diagonal terms are set).

The remainder of the example is almost entirely boilerplate \latex, which one might find in the source code of a journal article. In Lns.~5--8 the diagonal components of the Schwarzschild metric are specified. Line~9 is the final configuration command, which assigns $g_{\mu\nu}$ as the metric, and in doing so automatically constructs the metric inverse $g^{\mu\nu}$ and the associated connection $\Gamma^\sigma_{\mu\nu}$, both of which are essential to define the Riemann curvature tensor $R^\alpha{}_{\beta\mu\nu}$, the Ricci curvature tensor $R_{\beta\nu}$, the Ricci scalar $R$, the Einstein tensor $G_{\beta\nu}$, and the Kretschmann scalar $K$:
\begin{eqnarray}
    R^\alpha{}_{\beta\mu\nu} &=& \partial_\mu \Gamma^\alpha_{\beta\nu} - \partial_\nu \Gamma^\alpha_{\beta\mu} + \Gamma^\alpha_{\mu\gamma} \Gamma^\gamma_{\beta\nu} - \Gamma^\alpha_{\nu\sigma} \Gamma^\sigma_{\beta\mu} \\
    R_{\beta\nu} &=& R^\alpha{}_{\beta\alpha\nu} \\
    R &=& g^{\beta\nu} R_{\beta\nu} \\
    G_{\beta\nu} &=& R_{\beta\nu} - \frac{R}{2} g_{\beta\nu} \\
    K &=& R^{\alpha\beta\mu\nu} R_{\alpha\beta\mu\nu}. 
\end{eqnarray}

Upon parsing the contents of Fig.~\ref{fig:schwarzschild}, \nrpylatex generates symbolic (\sympy) expressions for all defined tensors and scalars using \nrpy notation as described in Sec.~\ref{sec:method}, and injects these expressions into the local namespace. So for example one can immediately access $g_{\theta\theta}$ from \verb!gDD[2][2]!, finding that indeed it stores $r^2$. Similarly, one can immediately confirm that each component of the Einstein tensor $G_{\mu\nu}$, stored in the nested list \verb!GDD[mu][nu]! is identically zero (this requires use of \sympy's simplification routine), and that the Kretschmann scalar (stored within the local variable \verb|K|) is equal to the known value for the Schwarzschild solution $K=48(GM/r)^2$. Thus we are able to confirm that \nrpylatex is correctly parsing these complex symbolic expressions, verifying that the Schwarzschild metric indeed solves Einstein's equations and that invariants take values consistent with the literature.

Next we move on to a workload in which the metric is only known numerically, and thus the metric and metric derivatives must be left in a general unspecified form. In this workload, the resulting expressions can be directly input into \nrpy, which outputs highly optimized \ccpp-code kernels that can adopt finite-difference representations to approximate partial derivatives.

\subsection{Numerical Workload Example: BSSN Hamiltonian Constraint for Vacuum Spacetimes with Unspecified Metric}
\label{ex:hamiltonian}

The 3+1 Baumgarte-Shapiro-Shibata-Nakamura (BSSN) formalism of general relativity~\cite{Baumgarte:1998te,Shibata:1995we} is perhaps the most widely used formulation within the numerical relativity community. Transcribing the BSSN equations into a hand-optimized \texttt{C/C++/Fortran} code can be a daunting task by itself, not to mention the debugging process. Here we present a \nrpylatex workflow aimed at generating a \sympy expression for the BSSN Hamiltonian constraint (i.e., the Hamiltonian constraint written in terms of BSSN variables) in vacuum\footnote{The \nrpy tutorial contains a Jupyter notebook for parsing the entire BSSN formalism in vacuum: \url{https://github.com/zachetienne/nrpytutorial}.}. The resulting expression can be plugged directly into \nrpy to generate highly optimized \ccpp-code kernels.

We start by defining the BSSN conformal 3-metric $\bar{\gamma}_{ij}$, which is related to the the physical 3-metric $\gamma_{ij}$ via
\begin{equation}
    \bar{\gamma}_{ij} = e^{-4\phi} \gamma_{ij}.
\end{equation}
Further the BSSN formalism decomposes the physical extrinsic curvature $K_{ij}$ into its trace $K$ and its trace-free part $\bar{A}_{ij}$:
\begin{equation}
    \bar{A}_{ij} = e^{-4\phi} \left(K_{ij} - \frac{1}{3} \gamma_{ij} K\right).
\end{equation}

Given these definitions, the Hamiltonian constraint is written (Eq.~46 of~\cite{Ruchlin:2017com})
\begin{equation}
    H = \frac{2}{3} K^2 - \bar{A}_{ij} \bar{A}^{ij} + e^{-4\phi} \left(\bar{R} - 8 \bar{D}^i \phi \bar{D}_i \phi - 8 \bar{D}^2 \phi\right),
\end{equation}
where $\bar{D}_i$ and $\bar{R}$ are the covariant derivative and Ricci scalar associated with $\bar{\gamma}_{ij}$, respectively.

The following example was written as a validation test against the same expression manually input into \nrpy (and independently validated against other trusted BSSN codes). As such, it adopts the \nrpy convention referring to the evolved ``conformal factor'' variable as ${\rm cf}$. \nrpy natively supports three options for conformal variable, but following e.g.,~\cite{Montero:2012yr,Marronetti:2007wz,Ruchlin:2017com} instead of evolving $\phi$ or $e^{-4\phi}$ directly in a numerical relativity code when constructing the spacetime, the variable ${\rm cf}=W=e^{-2\phi}$ is evolved. This variable is generally chosen as it is a smoother function near puncture black holes, resulting in lower truncation errors when evaluating numerical derivatives.

\begin{figure}[H]
\begin{boxed}
% vardef -diff_type=dD -metric gammabarDD @%% append _dD to each partial derivative@
% vardef -diff_type=dD -symmetry=sym01 AbarDD, RbarDD @%% symmetry [i][j] == [j][i]@
% parse \bar{R} = \bar{\gamma}^{ij} \bar{R}_{ij} \\ @%% internal calculation (no rendering)@
% vardef -diff_type=dD cf
% srepl "e^{-4\phi}" -> "\text{cf}^2" @%% syntactic string replacement@
% srepl -persist "\partial_<1> \phi" -> "\partial_<1> \text{cf} \frac{-1}{2 \text{cf}}"
% srepl "\bar{D}^2" -> "\bar{D}^i \bar{D}_i" @%% custom operator definition@
H = \frac{2}{3} K^2 - \bar{A}_{ij} \bar{A}^{ij} + e^{-4\phi} \left(\bar{R} - 8 \bar{D}^i \phi \bar{D}_i \phi - 8 \bar{D}^2 \phi\right)
\end{boxed}
\vspace*{-0.9\baselineskip}
\captionof{figure}{Numerical/\nrpy workload example: The Hamiltonian constraint of the BSSN formalism}
\label{fig:hamiltonian}
\end{figure}

In Fig.~\ref{fig:hamiltonian}, we demonstrate the procedure for constructing the Hamiltonian constraint in terms of BSSN variables using as input unspecified spacetime variables $W$, $K$, $\bar{A}_{ij}$, $\bar{\gamma}_{ij}$, and $\bar{R}_{ij}$, and any derivatives will be computed numerically (outside of \nrpylatex) from these given variables. By ``unspecified'', we mean that e.g., each component of the conformal 3-metric is left completely symbolic (e.g. $\gamma_{01}$ would be the \sympy symbol \verb!gammaDD01!, accessible from the local namespace variable \verb!gammaDD[0][1]!). Further, unlike the example of Sec.~\ref{ex:schwarzschild} (Ln.~1), we do not specify a particular coordinate system as this expression is covariant.

Line 1 defines an unspecified, 3-dimensional metric $\bar{\gamma}_{i j}$ and changes the derivative type to append a suffix \verb!_dD! instead of attempting to evaluate each derivative symbolically. The resulting symbolic expressions involving e.g., \verb!gammabarDD_dD[j][k][i]! (i.e., $\bar{\gamma}_{jk,i}$) can be passed into \nrpy, which generates \ccpp-code kernels that evaluate the derivatives using finite differences. Next (Ln.~2) the symmetric tensors \verb!AbarDD! and \verb!RbarDD! are defined in a similar way (i.e., they each appear in the final expression). As \verb!AbarDD! and \verb!RbarDD! are unspecified (i.e., we assume they are computed independently, and thus used here as input into the expression), their symmetry in the first two indices ``\verb!sym01!'' (i.e. $\bar{A}_{ij}=\bar{A}_{ji}$ and $\bar{R}_{ij}=\bar{R}_{ji}$) must be manually imposed. Then (Ln.~3) the scalar \verb!Rbar! is constructed by contracting \verb!RbarDD! with itself, i.e. $\bar{R}=\bar{\gamma}^{ij}\bar{R}_{ij}$.\footnote{Note, we could also use $\bar{R}=\bar{R}^i{}_i$ instead of $\bar{R}=\bar{\gamma}^{ij}\bar{R}_{ij}$ since \nrpylatex performs automatic index raising and lowering using the metric.}

Next, we apply a two-step procedure to replace every instance of $\phi$ with the chosen evolved ``conformal factor'' variable $\mathrm{cf}=W=e^{-2\phi}$. First, we define the conformal factor \verb!cf! (Ln.~4) and replace every instance of $e^{-4\phi}$ with $\mathrm{cf}^2$ (Ln.~5). Note that this and all \verb!srepl! replacements are done on the syntactic level, and are {\it not} string replacements. Second (Ln.~6), we replace each partial derivative of the form $\partial_{\langle 1\rangle} \phi$ for any index $\langle 1\rangle$ with the expression $(-1/2) \mathrm{cf}^{-1} \partial_{\langle 1\rangle} \mathrm{cf}$, as implied by the chain rule:
\begin{equation*}
    \partial_{\langle 1\rangle}\mathrm{cf}=\partial_{\langle 1\rangle}e^{-2\phi}=-2e^{-2\phi}\partial_{\langle 1\rangle}\phi\Rightarrow\partial_{\langle 1\rangle}\phi=-\frac{1}{2}\mathrm{cf}^{-1} \partial_{\langle 1\rangle} \mathrm{cf}.
\end{equation*}
Line 6 also enables the \verb!-persist! option, which ensures this replacement is made inside of each covariant derivative expansion.

Finally (Ln.~7), \verb!srepl! is used one more time to define a custom contraction operator $\bar{D}^2\mathrel{\mathop:}=\bar{D}^i \bar{D}_i$. \nrpylatex expands each covariant derivative using the metric $\bar{\gamma}_{ij}$ and the associated, internally generated connection $\bar{\Gamma}^i_{jk}$. All components of $\bar{\Gamma}^i_{jk}$ are available to the user in their local namespace, as well as the covariant derivatives (represented with the suffix \verb!_cd[UD]..[UD]!; the \verb!..! notation here denotes repetition to match the rank, or number of indices, of an indexed symbol). All this defined, the Hamiltonian constraint is immediately constructed in a form ready for \nrpy\ \ccpp-code kernel generation. 

\subsection{Example 3: \nrpylatex Exception Handling and Index Checking}
\label{ex:exceptionhandling}
\begin{figure}[H]
\vspace{-0.4cm}
\begin{iboxed}
`In [`!1!`]:` %load_ext nrpylatex.extension

`In [`!2!`]:` %%parse_latex
   ...: % vardef gUD, vD
   ...: v^a = g^a_b v_b
@TensorError@: illegal bound index 'b' in vU

`In [`!3!`]:` %%parse_latex
   ...: % vardef gUU
   ...: v^a = g^{cb} v_b
@TensorError@: unbalanced free index {'a', 'c'} in vU

`In [`!4!`]:` %parse_latex v^a = g^{a*} v_b
@ParseError@: v^a = g^{a*} v_b
                      ^
unexpected '*' at position 10

`In [`!5!`]:` %parse_latex T_{ab} = v_a w_b
@ParseError@: T_{ab} = v_a w_b
                         ^
cannot index undefined tensor 'wD' at position 32

`In [`!6!`]:` %parse_latex J^a = (4\pi k)^{-1} \nabla_b F^{ab}
@ParseError@: J^a = (4\pi k)^{-1} \nabla_b F^{ab}
                                ^
cannot generate covariant derivative without defined metric 'g'
\end{iboxed}
\vspace*{-\baselineskip}
\captionof{figure}{Examples of exception handling within an \ipython (interactive \python)/\jupyter session of \nrpylatex}
\label{fig:exceptionhandling}
\vspace{-0.4cm}
\end{figure}

Fig~\ref{fig:exceptionhandling} demonstrates \nrpylatex's robust system for index checking and generic exception handling, implemented using its own \verb!TensorError! and \verb!ParseError! algorithms, respectively. In short these algorithms report ambiguities in mathematical expressions, violations of Einstein notation, and problems at the syntactic level with error reporting that is human-friendly, both being descriptive and pointing to the exact location of the error in either the \nrpylatex configuration or \latex expressions.

\nrpylatex will raise its \verb!TensorError! exception upon violation of the Einstein summation convention. In Fig~\ref{fig:exceptionhandling}, Cell 1 we load the custom \ipython/\jupyter extension module \verb!nrpylatex.extension! using the built-in \verb!load_ext! magic command.\footnote{In a standard \python environment, we would instead import the \verb!parse_latex! function from the \verb!nrpylatex! module using the \verb!import! keyword.} In Cell 2 we attempt to parse the equation $v^a = g^a{}_b v_b$, which is invalid as Einstein notation requires that a repeated (``bound'') index may appear only once as a superscript and exactly once as a subscript in any given term. \nrpylatex thus raises a \verb!TensorError! exception for an `illegal bound index', which occurred since the index \verb!b! appeared twice in the \textit{down} position. Next, in Cell 3 we attempt to parse the equation $v^a = g^{cb} v_b$, proper Einstein summation notation requires a free index to appear in every term with the same (``up'' or ``down'') position and cannot be summed over in any term. \nrpylatex checks for this and raises a \verb!TensorError! exception for an `unbalanced free index', which occurred since index \verb!a! and index \verb!c! appeared only once on each side of the equation.

\nrpylatex will raise the custom \verb!ParseError! exception upon identifying a syntax error in the argument string of the \verb!parse_latex! function or magic command. In Cell 4, we attempt to parse the equation $v^a = g^{a*} v_b$, but \nrpylatex raised a \verb!ParseError! exception for an `unexpected' lexeme as
\nrpylatex did not recognize a valid index. \nrpylatex could also raise a \verb!ParseError! exception for an `expected' token if only that exact token would satisfy a production rule (e.g. \verb!\sqrt[0.2]{...}! would raise that exception since \nrpylatex would expect an integer for the root). In Cell 5, we attempt to parse the equation $T_{ab} = v_a w_b$, but \nrpylatex raises a \verb!ParseError! exception for indexing an `undefined' symbol \verb!wD!, which occurred because the indexed symbol \verb!wD! has not been defined (i.e., \nrpylatex did not find the indexed symbol \verb!wD! in the namespace). This exception can be addressed by defining \verb!wD! with the \verb!vardef! macro. In Cell 6 we attempt to parse the equation $J^a = (4\pi k)^{-1} \nabla_b F^{ab}$, but \nrpylatex raised a \verb!ParseError! exception since a covariant derivative cannot be generated without first defining a metric. For covariant derivatives without a diacritic, \nrpylatex searches the namespace for the metric \verb!gDD! (or \verb!gUU!) and did not find it, thus throwing the error. The Appendix describes more generally which metric is assumed to be associated with a given covariant derivative.

\section{Conclusions \& Future Work}
\label{sec:conclusions}

\nrpylatex provides a \latex interface for the free and open source \sympy CAS, and includes native support for Einstein notation. As \latex is widely used in the scientific community and the \nrpylatex infrastructure is completely free and open source, the learning curve for new users is minimized; the financial barrier to entry is eliminated; and anyone is free to modify and extend \nrpylatex (within the confines of its permissive 2-Clause BSD license). In addition to Einstein notation, \nrpylatex can, provided a metric, dynamically perform index raising and lowering, and automatically generate the metric inverse, determinant, and connection. Moreover, \nrpylatex expands covariant and Lie derivatives; and generates the Levi-Civita symbol (of arbitrary rank).

\nrpylatex's configuration consists of a command system modeled on the POSIX command syntax, aiming to reduce the learning curve. In brief, the command system defines variables, keywords, and ignore commands; assigns properties and attributes; and performs (syntactic) string replacement. To prevent rendering of \nrpylatex commands, we embed each command inside of a \latex comment; i.e., in text that follows a single percent symbol. This enables entire \nrpylatex workflows to be stored seamlessly within a scientific paper's \latex source code.

In summary, use cases for \nrpylatex range from symbolic manipulation of complex tensorial expressions in \jupyter notebooks, numerical code generation (e.g., when combined with its sister code \nrpy), and sharing of mathematical workflows directly within the \latex source code of a scientific paper.

In the future, we plan on expanding the integration with \nrpy to include reference metric support. This will enable one to choose from the broad class of curvilinear coordinate systems that \nrpy supports~\cite{Ruchlin:2017com}, to output covariant expressions in a particular coordinate system. In addition, we plan to incorporate native support for Post-Newtonian (PN) notation. In the current build of \nrpylatex, we are capable of parsing PN notation using the \verb!srepl! command to remap the PN notation for the dot/cross products of vectors to their definitions in Einstein notation. However, this approach requires too many \verb!srepl! replacements to be both robust and feasible. Instead, we are targeting native PN support for a future release of \nrpylatex. Finally, we plan on adding a direct pipeline for parsing \latex to optimized \ccpp-code using \nrpy.

\section*{Acknowledgments}
The authors would like to thank S.~R.~Brandt, R.~Haas, and D.~Chiang for useful discussions and suggestions during the preparation of \nrpylatex. Support for this work was provided by NSF awards OAC-2004311, PHY-1806596, as well as NASA awards ISFM-80NSSC18K0538 and TCAN-80NSSC18K1488.

\appendix
\section*{Appendix (\nrpylatex v1.1.1 Reference Manual)}
\label{app:blah}
\setcounter{section}{1}

This Appendix serves both to ensure completeness in the exposition of \nrpylatex features and to provide a quick reference guide for \nrpylatex.

\subsection{PARSE\_LATEX Function}
\begin{figure}[H]
\begin{tikzpicture}
\node[mybox](box){%
\begin{minipage}{0.95\textwidth}
\begin{titleboxed}
Parameters
----------
sentence : str
    input string (raw string preferred)
reset : bool, default=False
    reset current state by clearing namespace
verbose : bool, default=False
    verbose output and visible traceback
ignore_warning : bool, default=False
    ignore OverrideWarning
Returns
-------
tuple of str or sympy.core.expr.Expr
    tuple of each new variable (inserted into current stackframe) or
    single SymPy expression (sentence must also be an expression)
Raises
------
TensorError
    violation of the Einstein summation convention
ParseError
    invalid input string (related to SyntaxError)
Warns
-----
OverrideWarning
    a variable in the namespace is overridden
\end{titleboxed}
\end{minipage}
};
\node[fancytitle,right=10pt] at (box.north west){\verb!parse_latex()!};
\end{tikzpicture}
\vspace*{-\baselineskip}
\captionof{figure}{PARSE\_LATEX Function Reference}
\label{fig:parselatex}
\end{figure}

The \verb!parse_latex! function provides a frontend interface to \nrpylatex, accepting a \latex string, preferably a \python \textit{raw string} to avoid unwanted character escaping, and returning either a \sympy expression or a namespace dictionary containing every variable defined by the \verb!vardef! command and/or a \latex equation. Note, \nrpylatex injects every instantiated variable into the local namespace (i.e. the local scope of the \verb!parse_latex! function call) to avoid redundant variable assignment. Further, to eliminate constant redefining of common symbols/tensors, \nrpylatex automatically expands Levi-Civita symbols (of arbitrary rank), covariant derivatives, and Lie derivatives with their appropriate Einstein notation definitions using an operator inference system. In addition to parsing expressions/equations in Einstein notation, \nrpylatex also supports the \latex \texttt{align} environment for parsing multiple equations in a single \verb!parse_latex! function call. \nrpylatex also supports comma/semicolon notation for specifying partial/covariant derivatives, respectively. Finally, \nrpylatex provides an \ipython/\jupyter extension module \verb!nrpylatex.extension! that includes a \verb!parse_latex! (line and cell) magic command, which aliases to the \verb!%%latex! cell magic for displaying and the \verb!parse_latex! function for processing.

\subsection{PARSE Command}

\begin{figure}[H]
\begin{tikzpicture}
\node[mybox](box){%
\begin{minipage}{0.95\textwidth}
\begin{titleboxed}
USAGE
    parse EQUATION
EQUATION
    single LaTeX assignment + '\\'
\end{titleboxed}
\end{minipage}
};
\node[fancytitle,right=10pt] at (box.north west){\verb!parse!};
\end{tikzpicture}
\vspace*{-\baselineskip}
\captionof{figure}{PARSE Command Usage}
\label{fig:parse}
\end{figure}

The \verb!parse! command parses an equation without typesetting in a \verb!.tex! document or a Jupyter Notebook. Typically, we use the \verb!parse! command to perform a intermediate calculation that we deem necessary but trivial; e.g., a contraction to reduce rank.

\subsection{IGNORE Command}
\begin{figure}[H]
\begin{tikzpicture}
\node[mybox](box){%
\begin{minipage}{0.95\textwidth}
\begin{titleboxed}
USAGE
    ignore SUBSTRING, ...
SUBSTRING
    double-quoted string
\end{titleboxed}
\end{minipage}
};
\node[fancytitle,right=10pt] at (box.north west){\verb!ignore!};
\end{tikzpicture}
\vspace*{-\baselineskip}
\captionof{figure}{IGNORE Command Usage}
\label{fig:ignore}
\end{figure}

The \verb!ignore! command ignores a \latex command or substring during parsing, so that one can sanitize \latex typesetting for input into \nrpylatex. Note, the following formatting commands are ignored by default: \verb!\left!, \verb!\right!, \verb!{}!, and \verb!&!. Furthermore, the \verb!ignore! command and the empty replacement (\verb!% srepl "..." -> ""!) are interchangeable. Finally, we remark that an equation cannot be split across a line break, and hence we suggest appending a percent symbol \verb!%! to the end of a line break, creating a custom line break \verb!\\%!, and then removing that substring using the ignore command.

\subsection{VARDEF Command}
\begin{figure}[H]
\begin{tikzpicture}
\node[mybox](box){%
\begin{minipage}{0.95\textwidth}
\begin{titleboxed}
USAGE
    vardef [OPERAND ...] [OPTION] VARIABLE [DIMENSION], ...
OPERAND
    metric=METRIC
        desc: assign metric for index manipulation
        default: metric associated with diacritic (or lack thereof)
    weight=NUMBER
        desc: assign weight for Lie derivative generation
        default: 0
    diff_type=DIFF_TYPE
        desc: assign derivative type {symbolic | dD | dupD (upwind)}
        default: symbolic
    symmetry=SYMMETRY
        desc: assign (anti)symmetry
        default: nosym
        example(s):  sym01 -> [i][j] = [j][i], anti01 -> [i][j] = -[j][i]
OPTION
    const
        label variable type: constant
    kron
        label variable type: delta function
    metric
        label variable type: metric
    zero
        assign zero to each component
VARIABLE
    alphabetic string
    example(s): vU, gDD, alpha
DIMENSION
    variable dimension (array length)
    default: 3D
    example(s): vU::4D
\end{titleboxed}
\end{minipage}
};
\node[fancytitle,right=10pt] at (box.north west){\verb!vardef!};
\end{tikzpicture}
\vspace*{-\baselineskip}
\captionof{figure}{VARDEF Command Usage}
\label{fig:vardef}
\end{figure}

The \verb!vardef! command defines a variable, including optional assignment of a metric (for index raising/lowering), derivative type, tensor weight, and (anti)symmetry option. In addition to assigning attributes and properties, we can label the variable a scalar constant, Kronecker delta, or metric tensor. Note, the \verb!metric! option symmetrizes the variable and generates the metric inverse, determinant, and connection. In \nrpylatex, every variable in the namespace has an associated metric to automatically perform index raising and lowering on that variable, with the default metric following from the diacritic (or lack thereof) in the variable name (e.g., $v_i, \hat{v}_i, \bar{v}_i, \tilde{v}_i$). Note, \nrpylatex also supports diacritics on covariant derivatives (e.g., $D_i,\hat{D}_i,\ldots,\tilde{\nabla}_i$), which determines the metric used in the expansion of those covariant derivatives.

If we wanted to change the derivative type of a specific subexpression, we could use the \verb!vphantom! command, a pseudo in-line \latex comment for \nrpylatex, to prefix that subexpression with a derivative type. However, a priority system need be established between the derivative type of a variable and that specified by a \verb!vphantom! command. In general, \verb!vphantom! takes precedence over each variable in the subexpression with the exception of a variable explicitly assigned a symbolic derivative type. Finally, if \verb!v! and \verb!vU!, for example, are both in the namespace and we attempt to parse the expression $v^2$, the output from \nrpylatex will be \verb!vU[2]! (the third component of \verb!vU!). To resolve that indexing ambiguity between \verb!vU[2]! and \verb!v**2! (\verb!v! squared), we suggest using the notation \verb!v^{{2}}! since \verb!v^2! and \verb!v^{{2}}! are typeset identically in \latex. Similarly, if \verb!v! and \verb!vD! are both in the namespace and we are parsing the symbol $v_2$, we suggest replacing \verb!v_2! with \verb!\text{v_2}! using the \verb!srepl! macro to construct a compound symbol in \nrpylatex and preserve the \latex typesetting of $v_2$.

\subsection{ATTRIB Command}
\begin{figure}[H]
\begin{tikzpicture}
\node[mybox](box){%
\begin{minipage}{0.95\textwidth}
\begin{titleboxed}
USAGE
    attrib OPERAND VARIABLE
OPERAND
    coord=COORD
        desc: define coord (or coordinate system)
        example(s): [x, y, z], [r, \phi], default
    index=RANGE
        desc: override index range
        example(s): i::4D, [a-z]::2D
VARIABLE
    alphabetic string
    example(s): vU, gDD, alpha
\end{titleboxed}
\end{minipage}
};
\node[fancytitle,right=10pt] at (box.north west){\verb!attrib!};
\end{tikzpicture}
\vspace*{-\baselineskip}
\captionof{figure}{ATTRIB Command Usage}
\label{fig:attrib}
\end{figure}

The \verb!attrib! command overrides global properties of \nrpylatex, called \textbf{attributes}, e.~g., the coordinate system or index ranges (for looping or summation). To enable symbolic differentiation without imposing a specific coordinate system, \nrpylatex implements a generalized coordinate system $\{x_0,\ldots,x_n\}$ capable of dynamic dimension adjustment, with a default dimension of three. However, a specific coordinate system can be imposed on \nrpylatex using the \verb!attrib! command, allowing for coordinate indexing and symbolic differentiation with respect to a specific coordinate system. In addition to defining a coordinate system, the \verb!attrib! command can also override the default range of an index used in Einstein summation notation, making it useful for mixed dimension indexing.

\subsection{ASSIGN Command}
\begin{figure}[H]
\begin{tikzpicture}
\node[mybox](box){%
\begin{minipage}{0.95\textwidth}
\begin{titleboxed}
USAGE
    assign [OPERAND ...] [OPTION] VARIABLE, ...
OPERAND
    metric=VARIABLE
        desc: assign metric for index manipulation
        default: metric associated with diacritic (or lack thereof)
    weight=NUMBER
        desc: assign weight for Lie derivative generation
        default: 0
    diff_type=DIFF_TYPE
        desc: assign derivative type {symbolic | dD | dupD (upwind)}
        default: symbolic
    symmetry=SYMMETRY
        desc: assign (anti)symmetry
        default: nosym
        example(s):  sym01 -> [i][j] = [j][i], anti01 -> [i][j] = -[j][i]
OPTION
    metric
        label variable type: metric
VARIABLE
    alphabetic string
    example(s): vU, gDD, alpha
\end{titleboxed}
\end{minipage}
};
\node[fancytitle,right=10pt] at (box.north west){\verb!assign!};
\end{tikzpicture}
\vspace*{-\baselineskip}
\captionof{figure}{ASSIGN Command Usage}
\label{fig:assign}
\end{figure}

The \verb!assign! command assigns a \verb!vardef! option to an already existing variable in the namespace, which can be useful for assigning attributes to variables created through equation parsing.

\subsection{SREPL Command}
\begin{figure}[H]
\begin{tikzpicture}
\node[mybox](box){%
\begin{minipage}{0.95\textwidth}
\begin{titleboxed}
USAGE
    srepl [OPTION] RULE, ...
OPTION
    persist
        apply rule(s) to every subsequent input of the parse() function (internal or external)
RULE
    syntax: "..." -> "..."
    remark (1): whitespace ignored
    remark (2): substring can include a lexeme capture group
        syntax: <i> (single), <i..> (continuous) where i = 0, 1, 2, ...
\end{titleboxed}
\end{minipage}
};
\node[fancytitle,right=10pt] at (box.north west){\verb!srepl!};
\end{tikzpicture}
\vspace*{-\baselineskip}
\captionof{figure}{SREPL Command Usage}
\label{fig:srepl}
\end{figure}

The \verb!srepl! command performs string replacement, ignoring whitespace, with lexeme capture group support, allowing for dynamic variable renaming, expansion/replacement of subexpressions, custom operator definitions, custom superscript/subscript definitions (e.g., \verb!'!, \verb!+!, or \verb!-!), and much more. However, unlike their regular expression counterpart, \verb!srepl! capture groups can capture either a single lexeme or a sequence of lexemes\footnote{Consider the following pattern: \verb!\partial_t \phi = <1..> \\!. Now, suppose that instead of the \verb!srepl! capture group \verb!<1..>! we use a regex capture group \verb!([^\\]+)!. In that case, the string replacement would fail for substrings containing backslashes (a common occurrence in \latex).}. If we use a continuous capture group, then \verb!srepl! captures lexemes continuously until reaching the lexeme immediately following the capture group in the replacement pattern, e.g. \verb!x^{<1..>}! captures everything inside of the curly braces. Finally, the \verb!persist! option instructs \nrpylatex to perform (syntactic) string replacement on each \verb!parse_latex! function call, including those functions calls internal to \nrpylatex that generate \latex.

\renewcommand{\bibname}{References}
\printbibliography

\end{document}